\documentclass[12pt,a4paper]{article}
\usepackage{times}
\usepackage[preprint]{dfttrob}
\usepackage[numbers,sort&compress]{natbib}
\usepackage{latexuseful2e}
\usepackage{amsmath}
\usepackage{graphicx}
\DeclareTextFontCommand{\zapf}{\fontencoding{U}\fontfamily{pzd}\selectfont}

\dfttnum{DFTT 11/2002}

\begin{document}

\title{SUPERPOSITION EFFECT AND CLAN STRUCTURE\\
IN FORWARD-BACKWARD MULTIPLICITY CORRELATIONS}
\author{A. Giovannini and R. Ugoccioni\\
 \it Dipartimento di Fisica Teorica and I.N.F.N - Sez. di Torino\\
 \it Via P. Giuria 1, 10125 Torino, Italy}
\maketitle

\begin{abstract}
The main purpose of this paper is to discuss the link between
forward-backward multiplicity correlations properties and the shape of
the corresponding final charged particle multiplicity distribution in
various classes of events in different collisions.  It is shown that
the same mechanism which explains the shoulder effect and the $H_n$
vs.\ $n$ oscillations in charged particle multiplicity distributions,
i.e., the weighted superposition of different classes of events with
negative binomial properties, reproduces within experimental errors also
the forward-backward multiplicity correlation strength in \ee\
annihilation at LEP energy and allows interesting predictions for $pp$
collisions in the TeV energy region, to be tested at LHC, for instance
with the ALICE detector.  We limit ourselves at present to
study substructures properties in hadron-hadron collisions and \ee\
annihilation;  they are examined as ancillary examples in the
conviction that their understanding might be relevant also in other
more complex cases.
\end{abstract}

\section{Forward-backward correlations in \ee\ and $hh$ collisions.}

Forward-backward (FB) multiplicity correlations  have been studied
in hadron-hadron  collisions  and \ee\ annihilation
\cite{UA5:correlations,UA5:FB,UA5:rep,Benecke:FB,ISR:FB,NA22:FB,OPAL:FB,Delphi:FB,Tasso:FB}.
It has been found that these correlations are much stronger in hadron-hadron 
collisions than in \ee\ annihilation.

Let us start with  UA5 Collaboration results at CERN $p\bar p$ Collider
at 546 GeV c.m.\ energy \cite{UA5:FB,UA5:rep}.
It was found, by studying in each
event the  number of charged particles falling in the forward hemisphere,
$n_F$,  and in the backward    hemisphere, $n_B$, that the
relation between the average number of charged particles in the backward
hemisphere, $\nbar_B$,  and $n_F$ is  very well approximated by a
linear  one,
\begin{equation}
	\nbar_B (n_F) = a + b  n_F ,						\label{eq:nb_linear_nf}
\end{equation}
where $a$ is the intercept on the vertical axis and $b$ the slope of the
linear fit, i.e., the correlation strength: 
\begin{equation}
	b = \frac{\avg{(n_F-\nbar_F)(n_B-\nbar_B)}}
			{\left[ \avg{(n_F-\nbar_F)^2} \avg{(n_B-\nbar_B)^2}
			\right]^{1/2}}  .
													\label{eq:b_defined}
\end{equation}
Here the forward hemisphere corresponds to the region of the outgoing proton 
and  the backward hemisphere is  the symmetric region in the opposite 
direction.
In order to avoid correlations due to kinematical constraints like 
phase-space limits and energy momentum conservation  occurring at the border of 
the  rapidity range available in the collision, and short range correlations
produced from particle sources in more central rapidity intervals, the 
study has been performed in the pseudo-rapidity interval $1 < |\eta| < 4$. 
It has been found that $b$ parameter   is equal to $0.43 \pm 0.01$, a much 
larger value than that found at lower energy (ISR
and EHS) \cite{Benecke:FB,ISR:FB,NA22:FB},
e.g., $b = 0.156 \pm 0.013$ at 63 GeV c.m.\ energy. 
Assuming uncorrelated random emission of charged 
particles in the selected intervals of the pseudo-rapidity axis,  
the shape of the  $n_F$  multiplicity distribution  at 
fixed  full multiplicity $n = n_B+n_F$  is
binomial with probability
$p =1/2$ for a particle of the full sample, $n$, to fall in the backward
or forward hemisphere, and 
thus has variance $d^2_{n_F} (n)= p(1-p) n = n / 4$.
Since the experimental value of the dispersion of the full
distribution, $D_{n}$, in the above mentioned
pseudo-rapidity 
interval is $9.2 \pm 0.1$, and $\nbar = 15.8\pm 0.2$,  one has
\begin{equation}
   b = \frac{D^2_{n} - 4\avg{d^2_{n_F} (n)}}
						{D^2_{n} + 4\avg{d^2_{n_F} (n)}}
		= \frac{D^2_{n} - \nbar}{D^2_{n} + \nbar} = 0.69 ,
																								\label{eq:3}
\end{equation}
a much higher value than the experimental one ($b = 0.43 \pm 0.01$).
This fact led the Collaboration to assume that not particles but  particle
clusters of approximately the same size, $M$, are binomially distributed in the
two hemispheres and that the decay products of each cluster remain within
the same hemisphere.  Accordingly 
\begin{equation}
	4\avg{d^2_{n_F} (n)} = M n .
\end{equation}
But a reasonable agreement with experimental data is obtained by calculating
$M_{\text{eff}}$ for clusters of a mixture of sizes. One finds that
\begin{equation}
	b = \frac{D^2_{n}/\nbar - M_{\text{eff}}}{D^2_{n}/\nbar + M_{\text{eff}}}
		\qquad\text{with}\qquad
	M_{\text{eff}} = \bar M_{\text{cluster}} + 
				D^2_{\text{cluster}}/\bar M_{\text{cluster}}	.		\label{eq:M_eff}
\end{equation}
Here $\bar M_{\text{cluster}}$  and   $D^2_{\text{cluster}}$ 
are respectively the average charged
multiplicity  within a cluster and its dispersion.
An even better agreement is obtained by allowing a certain amount of
particle leakage from one hemisphere to the other.
In conclusion the correlation strength increases with energy and 
it is the result 
of binomially distributed clusters, of approximately 2.2  particles per 
cluster, in the two hemispheres. 

Forward-backward correlations have been studied also by OPAL
Collaboration \cite{OPAL:FB} in two-jet and three-jet events in \ee\
annihilation at LEP. The forward hemisphere is here chosen randomly
between the two defined by the plane perpendicular to the thrust axis;
although this definition is a bit misleading, a smaller effect than
that seen by UA5 collaboration is observed.  FB multiplicity
correlations are absent in the two separate samples of events but it
turns out that a correlation strength $b = 0.103 \pm 0.007$ is found
when they are superimposed, a result comparable with that found by the
DELPHI Collaboration ($b=0.118\pm 0.009$) \cite{Delphi:FB}.  By
comparing these results with that found at TASSO \cite{Tasso:FB},
e.g., $0.084\pm 0.016$ at 22 GeV,  we can conclude that in \ee\
annihilation the energy dependence of the correlation strength is
quite weak.

Our study is motivated by the just mentioned experimental facts and
by the finding that the superposition of weighted 
negative binomial (Pascal) multiplicity distributions (NB (Pascal) MD's),
each describing a different class of events, (soft and semi-hard
events in $pp$ collisions, 2- and 3-jet events
in \ee\ annihilation) explains quite well the characteristic features
of global  event properties of collisions in the GeV region,
like the shoulder effect in the total $n$-charged multiplicity distributions,
$P_n$,
and the $H_n$ vs.\ $n$ oscillations ($H_n$  is here the ratios of $n$-particle
factorial  moments, $F_n$, to the $n$-particle factorial cumulant moments,
$K_n$). The main purpose of this search is to show that FB multiplicity
correlations in $pp$ collisions and in \ee\ annihilation as well as
their different behaviour can be understood in terms of the same cause
which explained $P_n$ vs.\ $n$ and $H_n$ vs.\ $n$ general properties, 
i.e., the superposition of different substructures. 
In addition, the assumption
that each substructure (class of events) is described by a NB(Pascal)MD
allows sound quantitative predictions on parameters which
are not known from experimental data.
Coming to the correlation strength obtained in 
Eq.~(\ref{eq:3}), for instance,
if particles are independently produced and binomially distributed in
the two hemispheres, and the overall MD is a negative binomial with
characteristic parameters $\nbar$, the average charged multiplicity,
and $k$ (it is linked to the dispersion $D_n$ by the relation
$( D^2_n - \nbar)/\nbar^2 = 1/k$), one gets this simple Equation
\begin{equation}
	b = \frac{\nbar^2/k + \nbar - \nbar}{\nbar^2/k + \nbar + \nbar}
		= \frac{\nbar}{\nbar + 2k} .
										\label{eq:b_nbd_binom}
\end{equation}
A formula which can be applied also to individual  substructures
satisfying the requested conditions. 

The use of the NB(Pascal)MD in high energy phenomenology is  not indeed   
an arbitrary artifact but a consequence of our ignorance on  a problem 
(lack of experimental data and of explicit QCD calculations) and is quite
often  supported  by  fits with excellent chi squares  for charged
particles multiplicity distributions of the substructures appearing
 in  the total charged particle  multiplicity distributions 
through the above mentioned anomalies. 

The occurrence of the NB(Pascal)MD distribution in high energy
phenomenology   has been interpreted
since long time  as the by now quite well accepted idea that the
production process is a two-step process: to an initial phase in which
a certain number of sources, which have been called clans, are
independently emitted, it follows their decay into final particles.
All correlations among produced particles originated by the same
source  are exhausted within the same clan.  The average number of clans
$\Nbar$ is related to the standard NB parameters, $\nbar$ and $k$, by the 
following relation
\begin{equation}
            \Nbar = k \ln ( 1 + \nbar/ k).			\label{eq:7}
\end{equation}
Accordingly 
\begin{equation}   
            D^2_n = \nbar \exp( \Nbar/ k ) .		\label{eq:8}
\end{equation}

Each clan is either forward or backward. Produced particles by each clan
stay all in the same hemisphere where the clan is or some of them
may leak to the opposite hemisphere: the no-leakage  and leakage
cases will be discussed in Sec.~\ref{sec:3} under the  assumption  that
substructures in hadron-hadron collisions are described by NB(Pascal)MD
and therefore the concept of clan rather than the concept of cluster
should be used.

\section{Superposition of different classes of events and its influence
            on the correlation strength}

In this Section the effect of the superposition of different classes of events 
and its consequences on the strength $b$ of FB multiplicity
correlations will be discussed independently from the assumption that
charged particle MD's of the two classes of events are of NB type.
Results are therefore of general validity and  can be applied to
any pair of MD's describing experimental data in the two substructures
of the total MD.
Accordingly 
in the following the number 1 and 2 will indicate  the different
substructures (i.e., soft and semi-hard, or 2-jet and  3-jet events)
of the distribution in the two classes of collisions. $\alpha$ will be
the weight.

The joint distribution for $n_F$ and $n_B$ charged
particles is
\begin{equation}
	P_{\text{total}}(n_F,n_B) = \alpha P_1(n_F,n_B) + (1-\alpha) P_2(n_F,n_B) .
				\label{eq:superposition}
\end{equation}
In the following, the term `total' will be used used for quantities
referring to the superposition of the two components, and the
subscripts F and B stay as usual for `Forward' and `Backward'.

Since we have defined the F and B hemispheres in a symmetric way,
and since the collisions we are studying do not imply a difference
between F and B hemispheres, it is clear that for each class
$i=1,2$ of events, the joint MD's are symmetric in their arguments:
\begin{equation}
	P_i(n_F,n_B) = P_i(n_B,n_F) .
\end{equation}
in particular this implies that the average F multiplicity, 
$\bar n_{F,i}$, equals the average B multiplicity, $\bar n_{B,i}$, 
and both are equal to half the average multiplicity $\nbar_{i}$
in the two hemispheres, being $n_{i} = n_{F,i} + n_{B,i}$:
\begin{equation}
	\bar n_{F,i} = \bar n_{B,i} = \frac{1}{2} \nbar_{i} .
\end{equation}
We will also use below the equality of the variances:
\begin{equation}
	D^2_{n_F,i} \equiv \avg{n_{F,i}^2} - \nbar_{F,i}^2 =
	\avg{n_{B,i}^2}-\nbar_{B,i}^2 \equiv D^2_{n_B,i}.
\end{equation}
which together with Eq.~(\ref{eq:b_defined}) applied to each class of
events gives:
\begin{equation}
	\avg{(n_{F,i}-\nbar_{F,i})(n_{B,i}-\nbar_{B,i})} =
			b_i D^2_{n_F,i} = b_i D^2_{n_B,i};
\end{equation}
the relation with the total variance
then follows: 
\begin{equation}
	D^2_{n,i} = D^2_{n_F,i} + D^2_{n_B,i} + 
		2\avg{(n_{F,i}-\nbar_{F,i})(n_{B,i}-\nbar_{B,i})}
		= 2 (1+b_i) D^2_{n_F,i}  .
\end{equation}

Using Eq.~(\ref{eq:superposition}), we obtain the following relations
for the total average multiplicity and variance:
\begin{gather}
	\bar n_{F} = \bar n_{B} = \frac{1}{2} \nbar ;\\
	D^2_{n_F} = D^2_{n_B}.
\end{gather}
We start now by calculating the total forward average multiplicity 
$\bar n_{F}$:
\begin{equation}
	\bar n_{F} = \alpha \bar n_{F,1} + (1-\alpha)\bar n_{F,2}.
\end{equation}
We proceed to calculate the total forward variance $D^2_{n_F}$, using 
appropriately all the above mentioned relations:
\begin{equation}
	\begin{split}
		D^2_{n_F} &= \avg{n_{F}^2} - 
				  {\bar n_{F}}^2 \\
		&= \alpha \avg{n_{F,1}^2} + (1-\alpha) \avg{n_{F,2}^2} -
					\left[ \alpha \bar n_{F,1} + (1-\alpha) \bar n_{F,2}
			\right]^2\\
		&= \alpha D^2_{n_F,1} + (1-\alpha)D^2_{n_F,2} +
				\alpha (1-\alpha) \left( \bar n_{F,2} -  \bar n_{F,1}
			\right)^2\\
		&= \frac{\alpha D^2_{n,1}}{2(1+b_1)} +
			 \frac{(1-\alpha)D^2_{n,2}}{2(1+b_2)} +
				\frac{1}{4}\alpha (1-\alpha) 
							\left( \bar n_{2} -  \bar n_{1} \right)^2 .
	\end{split}
\end{equation}
The total covariance can be calculated as follows:
\begin{equation}
	\begin{split}
		 \avg{(n_F-\nbar_F)&(n_B-\nbar_B)} = \alpha \avg{n_{F,1} n_{B,1}} +
				(1-\alpha) \avg{n_{F,2} n_{B,2}} \\
			&\quad - \left[ \alpha \bar n_{F,1} + (1-\alpha) \bar n_{F,2}
			\right] \left[ \alpha \bar n_{B,1} + (1-\alpha) \bar n_{B,2}
			\right] \\
		&= \alpha \avg{(n_{F,1}-\nbar_{F,1})(n_{B,1}-\nbar_{B,1})} +
			(1-\alpha) \avg{(n_{F,2}-\nbar_{F,2})(n_{B,2}-\nbar_{B,2})} \\
				&\quad + \alpha (1-\alpha) \left( \bar n_{F,2} -  
							\bar n_{F,1} \right)^2\\
		&= \alpha b_1 D^2_{n_F,1} + (1-\alpha) b_2 D^2_{n_F,2} +
							\alpha (1-\alpha) \left( \bar n_{F,2} -  
							\bar n_{F,1} \right)^2\\
		&= \frac{\alpha D^2_{n,1}}{2(1+b_1)} b_1 +
				\frac{(1-\alpha)D^2_{n,2}}{2(1+b_2)} b_2 +
				\frac{1}{4}\alpha (1-\alpha) 
							\left( \bar n_{2} -  \bar n_{1} \right)^2  .
	\end{split}
\end{equation}

The total correlation strength $b$ for the weighted superposition of the two
classes of events can therefore be written as:
\begin{equation}
	b = \frac{\alpha b_1 {D^2_{n,1}}(1+b_2) +
				(1-\alpha) b_2 {D^2_{n,2}}(1+b_1) +
					\frac{1}{2}\alpha(1-\alpha)(\nbar_{2} - \nbar_{1})^2(1+b_1)(1+b_2)}
			{\alpha  {D^2_{n,1}}(1+b_2) +
				(1-\alpha)  {D^2_{n,2}}(1+b_1) +
					\frac{1}{2}\alpha(1-\alpha)(\nbar_{2} -
				\nbar_{1})^2(1+b_1)(1+b_2)}  .
																				\label{eq:b_total}
\end{equation}
It should be remarked that even if  the correlation strengths for events
of class 1 and 2, $b_1$ and $b_2$, 
separately vanish, the total strength does not: it           
depends on the difference in average multiplicity between
the events of the two classes, namely
\begin{equation}
	b_{12} = \frac{
					\frac{1}{2}\alpha(1-\alpha)(\nbar_{2} - \nbar_{1})^2}
			{\alpha  {D^2_{n,1}} +
				(1-\alpha)  {D^2_{n,2}} +
					\frac{1}{2}\alpha(1-\alpha)(\nbar_{2} - \nbar_{1})^2} .
																					\label{eq:b_12}
\end{equation}
Notice that in case
each component is of NBMD type, the above formulae can be rewritten
in terms of the standard NBMD parameters according to 
Eqs.~(\ref{eq:7}) and (\ref{eq:8}).

We conclude that  for $b_1=b_2=0$   forward-backward correlations are
different from zero: the superposition of events of different classes
generates a  certain amount of positive FB correlations.

It should be stressed  that up to now all results have been
obtained, as announced at the beginning of this Section,   independently
of any specific form of the charged particle MD's of the classes
of events 1 and 2  contributing to the total charged particle MD and that
Eq.~(\ref{eq:b_12})  gives  the amount of the superposition effect of
the substructures 1 and 2, $b_{12}$,  to the correlation strength with
$b_1=b_2=0$  in terms of   $\alpha, \nbar_1, D^2_1, \nbar_2, D^2_2$
parameters  only. Once the different classes of events have been isolated
and the above mentioned parameters measured, 
the estimate of $b_{12}$ can easily be done.
In \ee\ annihilation we are exactly in this situation  and our test
can be performed. 
It was shown by
the OPAL Collaboration \cite{OPAL:FB} that for the 2-jet and the
3-jet event samples, separately, there is no correlations. Still
there is correlation in the total sample, with $b = 0.103 \pm 0.007$.
Using a fit to OPAL data \cite{hqlett:2}
with similar conditions of the jet finder algorithm
($\alpha=0.463$, $\nbar_{1} = 18.4$, $D^2_{n,1} = 25.6$, 
$\nbar_{2} = 24.0$, $D^2_{n,2} = 44.6$) we obtain from
Eq.~(\ref{eq:b_12}) the value $b_{12} = 0.101$, in perfect agreement
with the data.

On the other hand, using UA5 two-components results \cite{Fug}
($\alpha=0.75$, $\nbar_{1} = 24.0$, $D^2_{n,1} = 106$, 
$\nbar_{2} = 47.6$, $D^2_{n,2} = 209$) in full phase-space
(thus including more correlations than present in the data,
but nothing better is available
in the actual phase-space range used by UA5) we obtain
$b_{12} = 0.28$, a number much lower than the value found
in the experiment (0.58).

We conclude that FB correlations generated by the superposition of events
of different classes  are enough to explain observed FB
correlations in \ee\ annihilation but not in $hh$ collisions,
where there exists a certain amount of correlation left within each
class of events which should be taken into account.
These results are a striking proof of the existence of the   superposition
effect, which was up to now  only  a guess, and of its relevance.

\section{Clan production and the correlation strength}\label{sec:3}
The next step in our approach is to calculate correlation strengths $b_1$
and $b_2$ in Eq.~(\ref{eq:b_total}) in the two different classes of events in
$pp$ collisions in  order to reproduce experimental data on $b$, 
which ---as pointed out at the end of the previous Section--- are not correctly
reproduced by the knowledge of $b_{12}$ only.
The success of the superposition mechanism of weighted  NB(Pascal)MD's 
for describing anomalies found in $pp$ collisions strongly suggest to 
proceed to calculate $b_1$ and $b_2$  by using NB properties.

In a naive approach to the problem one can try to apply
Eq.~(\ref{eq:b_nbd_binom}), which gives at 546 GeV c.m.\ energy $b =
0.78$: a much larger value than the experimental one (0.58).
We conclude that charged particles FB distribution is not compatible
with independent emission but is compatible with the production
in clusters. Within the framework of the NBMD, these will be
identified with clans.

As already mentioned, clans can be produced forward or backward and
may or may not leak particles to the opposite hemisphere.

\subsection{The no-leakage case}

First, we will treat the general case in which no assumption is made
about the MD of clans and of particles within a clan; then we will specialise
our results to the case of the NBMD (Poissonian clans, logarithmic MD
within a clan) obtaining a fair simplification of all formulae.
This treatment is valid for each component separately, but
the component index is dropped here to simplify the notation;
Eq.~(\ref{eq:b_total}) can be used to obtain the total correlation
strength.

Accordingly, we write the joint distribution in $n_F$,$n_B$
as a convolution over the number of produced clans:
\begin{equation}
	P(n_F,n_B) = \sum_{N_F,N_B} {\cal P}(N_F,N_B) p_F(n_F|N_F)
	p_B(n_B|N_B) ,
\end{equation}
where we have indicate with capital $N$ the number of clans and
with ${\cal P}$ the joint distribution for $N_F$ clans forward and
$N_B$ clans backward. 
Here $p_F(n|N)$ is the forward particle multiplicity
distribution conditional on the number of forward clans, which, by arguments
of symmetry, is the same distribution as $p_B(n|N)$.

The symmetry of the reaction and of the hemispheres definition imply
some conditions on ${\cal P}$, namely
that the average number of F and B clans at fixed full number of clans $N$
are equal, and similarly for the corresponding variances:
\begin{gather}
	\Nbar_F(N) = \Nbar_B(N) = N/2 ; \\
	d^2_{N_F}(N) = d^2_{N_B}(N) = \avg{N_F^2(N)} - N^2/4.
\end{gather}

If we now indicate with $q(n)$ the MD within one clan, i.e., we
write:
\begin{equation}
	p_F(n|N) = \underset{\{\sum_i n_i=n\}}{\sum_{n_1}\cdots\sum_{n_N} }
			\,q(n_1)\cdots q(n_N) ;
								\label{eq:breakintoclans}
\end{equation}
then it is straightforward to show that the average value and the
variance of $n$ at fixed $N$ equal $N\nc$ and $ND^2_c$ respectively.
Using these results, it can be shown that
	\begin{equation}
		\avg{n_F n_B} = \avg{N_F N_B}\nc^2  = 
			\left( \frac{1}{4}\avg{N^2} - \avg{d^2_{N_F}(N)} \right)\nc^2 
	\end{equation}
and
	\begin{equation}
		\avg{n_F^2} = \avg{N_F(N)} D^2_c + \avg{N_F^2(N)} \nc^2 =
			\frac{1}{2}\Nbar D^2_c  + \nc^2\left( \avg{d^2_{N_F}(N)} + 
							\frac{1}{4}\avg{N^2}  \right) .
	\end{equation}

Then, without making any additional hypothesis on the
clan distributions used, we can calculate the variances
in the particle multiplicity:
\begin{equation}
	\avg{(n_F-\nbar_F)^2} =
	 \avg{(n_B-\nbar_B)^2} = 
			\frac{1}{2} \Nbar D^2_c + \nc^2\left[ \frac{1}{4} D^2_N + 
							\avg{d^2_{N_F}(N)} \right] ,
			\label{eq:var_noleakage}
\end{equation}
where $\Nbar$ is the average full number of clans and 
$D^2_N = \avg{N^2} - \Nbar^2$
is the variance of the full clan multiplicity distribution;
$d^2_{N_F}(N)$ is the variance in the distribution 
of the number of forward
clans given that the total number of clans is $N$, and 
$\avg{d^2_{N_F}(N)}$ is its average over $N$;
$\nc$ is the full average number of particles per clan, and $D^2_c$ is the
corresponding variance.
The covariance is
\begin{equation}
	\avg{(n_F-\nbar_F)(n_B-\nbar_B)} = 
		 \nc^2\left[ \frac{1}{4} D^2_N - \avg{d^2_{N_F}(N)} \right]  .
			\label{eq:covar_noleakage}
\end{equation}
The final result is thus:
\begin{equation}
	b = \frac{D^2_N - 4\avg{d^2_{N_F}(N)}}{D^2_N + 4\avg{d^2_{N_F}(N)} + 
				2\Nbar D^2_c / \nc^2} .
													\label{eq:b_noleakage}
\end{equation}
This result can be expressed in terms of quantities referring to the
full MD: since
\begin{equation}
	D^2_{n} = D^2_c \Nbar + D^2_N \nc^2  ,
\end{equation}
we obtain
\begin{equation}
	b = \frac{D^2_{n}/\nbar - D^2_c/\nc - 4\nc\avg{d^2_{N_F}(N)}/\Nbar}
			{D^2_{n}/\nbar + D^2_c/\nc + 4\nc\avg{d^2_{N_F}(N)}/\Nbar} .
\end{equation}
From the above formula, assuming binomially distributed clans, 
one can deduce Eq.~(\ref{eq:M_eff}).

The last formulae are of general validity, but now we specialise it
to the NBMD case, where
the  overall clan multiplicity distribution is Poissonian and
thus one finds $D^2_N= \Nbar$, and where
clans do not talk to each other in this framework, thus
the forward distribution at fixed total number of clans
is binomial, $\avg{d^2_{N_F}(N)} =\Nbar/4$;
putting these features together in Eq.~(\ref{eq:b_noleakage}) 
we obtain:
\begin{equation}
	b = \frac{\Nbar - \Nbar}{\Nbar + \Nbar + 
				2\Nbar D^2_c / \nc^2} = 0 .
\end{equation}
It follows the theorem:  
If particles are grouped in Poisson distributed clans;
if each clan falls
in the forward or backward hemisphere with the same probability
independently of the other clans;
if there is no leakage of particles from one hemisphere to the other:
then no forward-backward correlation exist.

We conclude that in order to have FB correlation  in the
framework of the clan interpretation of the NBMD it is necessary
to allow clans to leak particles from one hemisphere to the other,
i.e., clans must extend rather far in rapidity.

\subsection{The extension to the leakage case}

We return now to the general treatment.  We start by writing the joint
distribution as a convolution over the number of produced clans and
over the partitions of forward and backward produced particles among
the clans:
\begin{equation}
	P(n_F,n_B) = \sum_{N_F,N_B} {\cal P}(N_F,N_B) 
				\mathop{\sum_{m_F'+m_F'' = n_F}}_{m_B'+m_B'' = n_B}
				p_F(m_F',m_B'|N_F) p_B(m_F'',m_B''|N_B)  ,
																				\label{eq:mainmodel}
\end{equation}
where $p_i(m_F',m_B'|N_i)$ is the joint probability of producing
$m_F$ F-particles and $m_B$ B-particles from $N_i$ clans in the $i$
hemisphere ($i=F,B$).
Notice that the summations are constrained to $n_F$ F-particles and
$n_B$ B-particles, respectively.
The usual symmetry argument imply that
\begin{equation}
	p_F(n,m|N) = p_B(m,n|N) .
\end{equation}

It is now straightforward to show that it is still true that
$\nbar_F = \Nbar \nc/2$.
Similarly to Eq.~(\ref{eq:breakintoclans}), we write now the
decomposition
\begin{equation}
	p_F(m_F,m_B|N_F) = 
    \mathop{\sum_{\sum_i \mu_{F,i} = m_F}}_{\sum_i \mu_{B,i} = m_B}
			q_F(\mu_{F,1},\mu_{B,1}) \dots q_F(\mu_{F,N_F},\mu_{B,N_F}) ,
\end{equation}
where $q_F(\mu_{F},\mu_{B})$ is the joint probability of producing,
within one clan, $\mu_F$ F-particles and $\mu_B$ B-particles.
Of course one has:
\begin{equation}
	\sum_{\mu_F+\mu_B = n_c} q_F(\mu_{F},\mu_{B}) = q(n_c) .
\end{equation}
Define now
\begin{gather}
	p\nc \equiv \sum \mu_F q_F(\mu_{F},\mu_{B}) = 
							\sum \mu_B q_B(\mu_{F},\mu_{B}),\\
	q\nc \equiv \sum \mu_B q_F(\mu_{F},\mu_{B}) =
						  \sum \mu_F q_B(\mu_{F},\mu_{B}) ,
\end{gather}
where the second equality in each row has been obtained from the usual symmetry
arguments, which also imply $p+q=1$.
The just defined parameter $p$ controls the leakage from one hemisphere to the
other: $p=1$ means that no particle leaks, while $0.5 \leq p < 1$
indicates leakage ($p$ cannot be smaller than 0.5, or else the
clan is classified in the wrong hemisphere).

An important role
in the final formula will be played by the covariance, $\gamma$, of
$\mu_{F}$ forward and $\mu_{B}$ backward particles 
within a clan:
\begin{equation}
	\gamma \equiv \avg{ (\mu_{F} - \bar\mu_{F}) (\mu_{B} - \bar\mu_{B})
	}  = \sum \mu_F \mu_B q_F(\mu_{F},\mu_{B}) - pq\nc .
\end{equation}
In general, this quantity cannot be expressed in terms of $\nc$
or $D^2_c$ unless some explicit distribution for the forward
distribution at fixed number of particles per clan is assumed;
for example, when particles within one clan are F-B
binomially distributed, then $\gamma = (D^2_c-\nc)pq$.

One then finds for the variance
\begin{equation}
	\avg{(n_F-\nbar_F)^2} =
	 \avg{(n_B-\nbar_B)^2} = 
			\frac{1}{2}\Nbar D_c^2 - \Nbar\gamma
	   +  \nc^2 \left[\frac{1}{4} D^2_N  + \avg{d^2_{N_F}(N)}(p-q)^2\right]
\end{equation}
---compare with Eq.~(\ref{eq:var_noleakage})--- and for the covariance
\begin{equation}
	\avg{(n_F-\nbar_F)(n_B-\nbar_B)} = 
			\Nbar\gamma + \nc^2\left[ \frac{1}{4}  D^2_N  - 
					 \avg{d^2_{N_F}(N)} (p-q)^2 \right]
\end{equation}
---compare with Eq.~(\ref{eq:covar_noleakage}).

The final general result for clans, for each component, is thus
\begin{equation}
	\begin{split}
	b &= \frac{D^2_N - 4\avg{d^2_{N_F}(N)}(p-q)^2 + 4\Nbar\gamma/\nc^2}
			{D^2_N + 4\avg{d^2_{N_F}(N)}(p-q)^2 - 4\Nbar\gamma/\nc^2 + 
					2\Nbar D^2_c/\nc^2}\\
		&= \frac{D^2_n/\nbar - D^2_c/\nc -
	4\avg{d^2_{N_F}(N)}(p-q)^2\nc/\Nbar + 4\gamma/\nc }{
	D^2_n/\nbar  + D^2_c/\nc +
	4\avg{d^2_{N_F}(N)}(p-q)^2\nc/\Nbar - 4\gamma/\nc }  .
	\end{split}
\end{equation}
For NBMD clans, i.e., Poissonian ($D^2_N = \Nbar$)
and independent ($4\avg{d^2_{N_F}(N)} = \Nbar$) clans, one has inside
a clan a logarithmic distribution, for which
\begin{equation}
	D^2_c = - \nc^2\left[ \frac{\log(1-b')}{b'} + 1\right]
\end{equation}
and
\begin{equation}
	\nc = \frac{b'}{(b'-1)\log(1-b')} ,
\end{equation}
with
\begin{equation}
	b' = \frac{\nbar}{\nbar+k}.
\end{equation}
The correlation strength can then be written as
\begin{equation}
	b = \frac{1 - (p-q)^2 + 4\gamma/\nc^2}
			{-1 + (p-q)^2 - 4\gamma/\nc^2 - 2\log(1-b')/b'} .
																	\label{eq:b_leakage}
\end{equation}
As anticipated, the correlation strength is no longer zero.
Notice that the no-leakage case can be re-obtained by setting
$p=1$ and consequently $q=0, \gamma = 0$. Recall that
$\gamma$ is the covariance between the forward and
backward multiplicities in one clan, so in Eq.~(\ref{eq:b_leakage}) only
parameters related to the within-clan distributions appear.
If one were to assume that particles within a clan are
independently distributed in the two hemispheres, so
that $\gamma = (D^2_c-\nc)pq$, then one arrives at a very
simple formula:
\begin{equation}
	b = \frac{2b'pq}{1-2b'pq} = \frac{-1+p}{1-p-\frac{1}{2b'p}} .
									\label{eq:b_leakage_binom}
\end{equation}
Notice that when $p=1/2$ we recover the expression for 
the NBMD with binomially distributed particles,
Eq.~(\ref{eq:b_nbd_binom}).
For $b' \to 1$, $b$ turns out to depend on the $p$ parameter only,
a fact which will be very useful in the next Section.

\section{Energy dependence of the correlation strength}  

\begin{figure}
  \begin{center}
  \mbox{\includegraphics[width=0.5\textwidth]{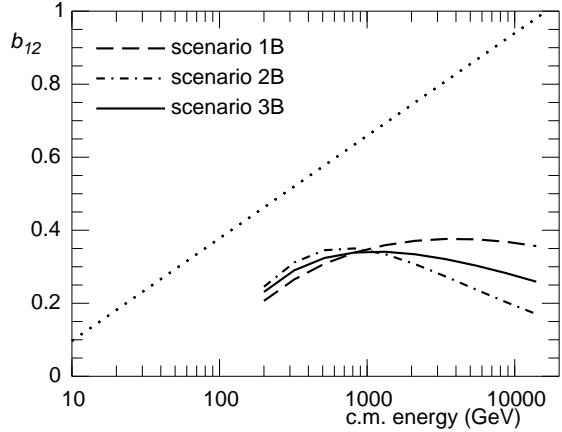}}
  \end{center}
  \caption{Predictions for the correlation coefficient for the
  superposition of two components in $p\bar p$ collisions
	in the case in which each
	component by itself presents no correlations;
	prediction are  given
	in the three scenarios in full phase-space as function
	of the c.m.\ energy.
	The dotted line is a fit to 
	experimental values \cite{UA5:correlations}.}\label{fig:b12}
  \end{figure}

In Fig.~\ref{fig:b12}, the behaviour of  $b_{12}$
as a function of c.m.\ energy, as suggested 
by  Eq.~(\ref{eq:b_12}), is shown in the framework of 
the three scenarios proposed 
by the present Authors in Ref.~\cite{combo:prd}.
The scenarios are based on the weighted superposition of soft (without
mini-jets) and semi-hard (with mini-jets) events in $p\bar p$
collisions, each class of events being described by a NB(Pascal)MD with
different characteristic parameters $\nbar$ and $k$.
In the first scenario KNO scaling is assumed for both components; in
scenario 2 KNO scaling is strongly violated for the semi-hard
component and satisfied for the soft one; the third scenario is a QCD
inspired scenario and its predictions turn out to be in general
intermediate between the previous two.
In addition, two alternative c.m.\ energy dependences have been
proposed for the semi-hard component: the first one is a consequence
of the UA1 analysis on mini-jets and leads to
\begin{subequations}
\renewcommand{\theequation}{\theparentequation \Alph{equation}}
\begin{equation}
	\nbar_2(\sqrt{s}) \simeq 2 \nbar_1(\sqrt{s}) ;
						\label{eq:A}
\end{equation}
the second one postulates that $\nbar_2$ increases more rapidly with
c.m.\ energy (it takes into account an eventual high particle density
production in the central rapidity region) and correct the previous
equation as follows
\begin{equation}
	\nbar_2(\sqrt{s}) \simeq 2 \nbar_1(\sqrt{s})+ c'\ln^2(\sqrt{s}) ;
						\label{eq:B}
\end{equation}
\end{subequations}
the estimate of $c'$ from existing fits is $\approx 0.1$.
The general trend of $\nbar$ as described by Eq.~(\ref{eq:B}) seems to be
favoured in the present approach.

It should be noticed that the superposition effect alone  does not
reproduce the logarithmic energy dependence of the correlation 
strength $b$ in the GeV and TeV regions. $b_{12}$ does not depend on $p_1$
and $p_2$ parameters, it is an increasing function of c.m.\ energy in
the GeV region hardly distinguishable in the three scenarios.
At 900 GeV c.m.\ energy, $b_{12}$ general trends overlap, then they
start to decrease smoothly and to differentiate their behaviour in the
TeV region.
Accordingly, in all scenarios Eq.~(\ref{eq:b_total}), and not
Eq.~(\ref{eq:b_12}), should be used in order to get the correct $b$ behaviour.
The role of $b_{12}$ has been shown to be fundamental in understanding
FB correlations in \ee\ annihilation  when
particle population within each clan are quite small ($\approx$ 1-2) and 
particle leakage from one hemisphere to the other quite an exceptional
fact (remember that here $b_1 \approx b_2 \approx 0$).
In proton-proton collisions the decrease of $b_{12}$ with c.m.\ energy
goes together with the onset of a much larger leakage activity from
clans with large particle population, characteristic in particular of
semi-hard events  in scenarios 2 and 3. 
It is interesting to remark indeed in
Fig.~\ref{fig:b12} that the decrease of $b_{12}$ in the TeV region
is more pronounced in these two
scenarios than in scenario 1 and in scenario 2 with respect to scenario 3,
i.e., in general when clans with larger number of particles are produced
and leakage effect is expected to be more important.

\begin{figure}
  \begin{center}
  \mbox{\includegraphics[width=\textwidth]{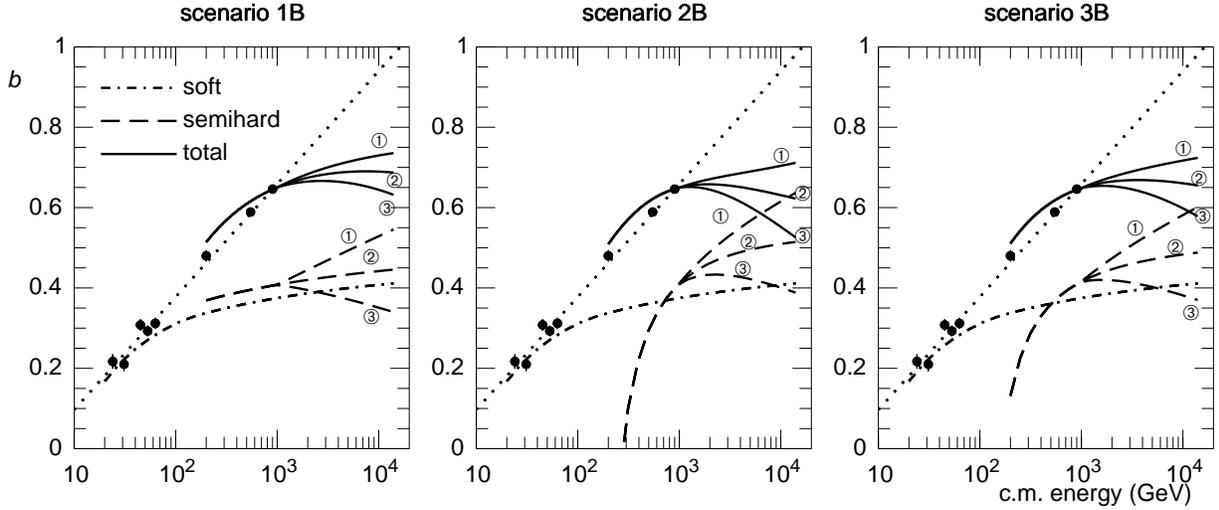}}
  \end{center}
  \caption{Predictions for the correlation coefficients 
	for each component (soft and semi-hard) 
	and for the total distribution in $p\bar p$ collisions.
	For each scenario, three
	cases are illustrated, corresponding to the three numbered branches: 
  leakage increasing with $\sqrt{s}$ (upper branch, \zapf{\char'300}),
	constant leakage (middle branch, \zapf{\char'301}) and
	leakage decreasing with $\sqrt{s}$ (lower branch, \zapf{\char'302}).
	Leakage for the soft component is assumed constant at all energies.
	The dotted line is a fit to 
	experimental values \cite{UA5:correlations}.}\label{fig:b}
  \end{figure}

Figure~\ref{fig:b} summarises  our findings on the c.m.\ energy dependence of
the FB charged particle  multiplicity correlation strength parameter $b$
both for the soft and semi-hard  components individually 
and  for their superposition
in the GeV and TeV regions up to 14 TeV  for the three above mentioned
scenarios.
By assuming that in $pp$ collisions in the ISR energy range  the
semi-hard component is negligible with respect to the soft one
and by using the experimental $b$ value at 63 GeV  
in Eq.~(\ref{eq:b_leakage_binom}),
the parameter $p_1$ controlling the leakage of particles emitted by
clans in one hemisphere to the opposite one  can be determined:
we find $p_1=0.78$, i.e., on the average 22\% of particle population
within  a clan are expected to leak in the opposite hemisphere. 
Since the average number of particles
per clan for the soft component goes from  $\approx 2$ at 63 GeV up 
to $\approx 2.44$  at 
900 GeV (clans are almost of the same size) it is quite reasonable
to conclude that  $p_1$  is approximately  energy independent in the
region.

By using the $p_1$ value  for the soft component at 546 GeV in
Eq.~(\ref{eq:b_leakage_binom}),
the value of the parameter $p_2$ controlling leakage effect in the  semi-hard
component  can also be determined: it is found that $p_2=0.77$, a pretty
close value to $p_1$. Clan structure analysis can help in understanding this 
result: average clan size goes from 1.64 at 200 GeV up to 2.63 at 900 GeV
for the semi-hard component. A very small difference with respect to 
the clan size for the soft component at the same c.m.\ energy.
In addition,
since the increase  of clan size for the semi-hard component from 200 GeV up
to 900 GeV  is indeed quite small, the  $p_2$ value 0.77 can be considered  
approximately also constant in the GeV energy range. 

With the just mentioned  constant values of $p_1$ and $p_2$
parameters, and assuming the c.m.\ energy dependence of the average charged
particle multiplicity $\nbar$ corrected by a $\ln^2 s$ term as
indicated in Eq.~(\ref{eq:B}), 
the general trend of the $b$ energy dependence  from ISR up to top 
$p\bar p$ CERN Collider energy  obtained by superimposing soft and semi-hard 
components effects is correctly reproduced. It agrees in particular with the
phenomenological fit proposed by different Collaborations 
\cite{UA5:correlations,NA22:FB} in the full
range, i.e., $b= -0.019 + 0.061 \ln s$ (dotted line in
Figs.~\ref{fig:b12} and \ref{fig:b}).

The just mentioned assumptions on $p_1$, $p_2$ and $\nbar$ are extended to the
TeV region.  At 900 GeV  a clear bending in the behaviour of the total strength
is visible in all three scenarios (full line \zapf{\char'301} in
Fig.~\ref{fig:b}) 
and is not compatible 
with the logarithmic increase of $b$ in the GeV energy range.
In view of this result, should one question  the validity of our assumptions 
on the constant behaviour  of  $p_1$ and $p_2$ leakage parameters for the
soft  and semi-hard components in the TeV region? Lack of sound experimental 
data prevent us from making sharp statements on the problem. A possible 
insight  comes again from clan structure analysis in the new energy domain.

The average number of particle per clan for the soft component is growing 
from 2.63 at 900 GeV up to 2.98 at 14 TeV (and from 2 at  63 GeV).
Therefore
constant $p_1$ seems  a quite well founded assumption  (it should be 
pointed out that variations of one-two  per cent in the $p_1$  and $p_2$ values
as will be the case by increasing  the average number of particles per clan
of one unit do not change  the overall scenario we are discussing).

The same conclusion can be drawn  in the TeV region for the 
semi-hard component
in scenario 1 where the average number of particles per clans goes from
2.63 at 900 GeV to 3.28 at 14 TeV and which shares with the soft component
KNO scaling properties. But it is not true in general when KNO scaling
violations are expected to occur: the average number of particle per clan
goes for instance in scenario 2 from 2.63 at 900 GeV up to 7.36 at 14 TeV:
one should see more particle leakage from clans 
and accordingly a decreasing $p_2$ value as c.m.\ energy
increases. This request can be taken into account  
as shown in Fig.~\ref{fig:b}:
$b$ bending becomes even higher assuming  that $p_2$ is for instance a 
logarithmic increasing function of c.m.\ energy in the TeV region and the
corresponding leakage effect decreasing (curves \zapf{\char'302} in the
figure).
The opposite situation  occurs in the three scenarios  and in particular in
scenario 2 and 3 assuming that $p_2$ is a logarithmically decreasing function
of the c.m.\ energy (curves \zapf{\char'300} in the figure)

A similar global $b$ behaviour is found indeed
in the three scenarios  when  corresponding
$D^2_n / \nbar^2$  values  are approximately the same.
Larger differences
appear when the increase of $\nbar$ with  c.m.\ energy  is not compensated by
a comparable decreasing of $k$ parameter.  It is also clear that
asymptotically the above mentioned ratio will depend on $k$ parameter only.

The bending of the total FB multiplicity correlation strength in
Fig.~\ref{fig:b}, in view of Eq.~(\ref{eq:b_leakage_binom}), is a
natural consequence of the fact that the quantity $\nbar/(\nbar + k)$
goes to 1 for increasing energy.  This consideration notwithstanding,
it is shown that an increasing leakage effect leads to an increase of
$b$ towards its maximum value $b=1$.  The impression is that in order
to get a quick saturation of $b$, larger leakage and clans with higher
population densities are needed: this could be a signal of a possible
onset of a third class of hard events, harder than soft and semi-hard
events discussed in this paper, producing the two requested effects.

Although our approach is limited to full phase-space only and therefore
kinematical constraints at the border of the allowed rapidity range
as well as short range correlations effects  from central rapidity
intervals  might influence  our claims, it seems quite reasonable
to say that the bending effect should be expected in the TeV
region. The deep  connection  between FB charged particle
multiplicity correlation strength $b$, leakage effects and clan structure
analysis in the framework of the superposition mechanism of 
different classes of events suggested by the present approach in order to 
try to understand and regulate $b$ bending  should also  
be tested in future experiment at LHC with the Alice detector.
Our results on $b$ bending effect should be compared with other
predictions based on different models \cite{jdd:FB,Carruthers:FB}.

Another aspect of the problem should be examined for completeness:
is the linear relations  between $\nbar_B(n_F)$ and $n_F$,
shown in Eq.~(\ref{eq:nb_linear_nf}),
affected by the assumptions which have been made in order to determine
the energy dependence of the total FB multiplicity 
correlation strength $b$?

\begin{figure}
  \begin{center}
  \mbox{\includegraphics[width=\textwidth]{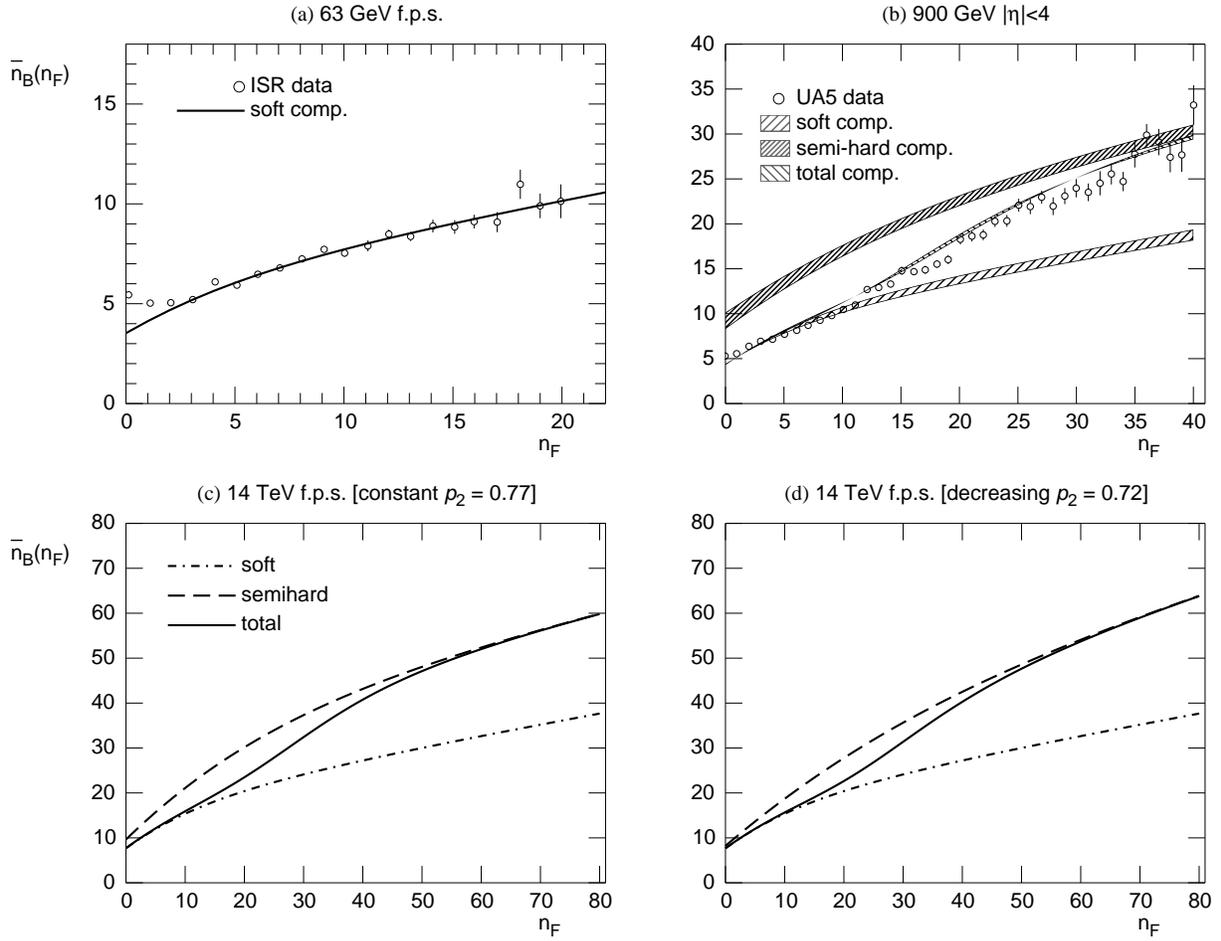}}
  \end{center}
  \caption{Results of our model for $\nbar_B(n_F)$ vs.\ $n_F$ compared 
	to experimental data \cite{ISR:FB,UA5:correlations}
	in full phase-space at 63 GeV (a) and in the
	pseudo-rapidity interval $|\eta|<4$ at 900 GeV (b).
	Theoretical predictions at 14 TeV in f.p.s. for constant $p_2$
	(c) and $p_2$ decreasing logarithmically with energy (d) are 
	also shown.}\label{fig:MC}
  \end{figure}

Following the von Bahr-Ekspong theorem \cite{Ekspong:theorem}, 
in fact, the linear behaviour of 
$\nbar_B(n_F)$ vs.\ $n_F$, the binomial distribution of particles in the
forward and backward hemispheres  and the occurrence of  the NBMD
are not independent statements. They are strongly linked: the validity
of any two  of them implies the validity of the third.
It is clear that the linearity of the relation is violated in the
soft and semi-hard components separately in our approach: in fact, it should
be pointed out  that in the first step  of the production process clans,
not particles, are binomially distributed in the two hemispheres with
$p=1/2$, and that in the second step each clan generates particles, binomially
distributed in the two hemispheres  but with $p_1$ and $p_2$ different from 1/2.
Accordingly, the final particles are not binomially distributed in the 
two hemispheres. Being the multiplicity distribution of the two
classes of events  a NBMD, the linearity of FB multiplicity 
correlations in each substructure cannot be exact. 
An example of a mild violation of linearity in the soft component
is shown 
in Fig.~\ref{fig:MC}(a) where  $\nbar_B(n_F)$ is plotted vs.\ $n_F$
at ISR  energies (63 GeV).
In addition, the extremely good fit of experimental data obtained by using
a single NB(Pascal)MD for describing this component 
which in this case represents the total MD
(at this energy the
semi-hard component has been assumed to be negligible) is an indirect
confirmation of the validity of our argument.
Of course by going to higher c.m.\ energies in $pp$ collisions one
expects  that the linearity violation occur not only in the soft
component but also in the semi-hard one.
The linearity violation in each separate components is shown
in Fig.~\ref{fig:MC}(b) 
at 900 GeV c.m energy: parameters from the two fits proposed in
\cite{Fug} have been used; their values determine the borders of the bands
shown in the figure;
the $p_2$ parameter has been taken equal to 0.77.

It should be pointed out that since the total multiplicity
distribution resulting from the weighted superposition of events of
the two separate substructures (with and without mini-jets) is not of
NB type, and in view of the lack of binomial
structure in the final charged particles MD
in the two hemispheres, 
the linear behaviour of $\nbar_B(n_F)$ vs.\ $n_F$ can in
principle be restored for the total MD in
agreement with the von Bahr-Ekspong theorem.

In Fig.~\ref{fig:MC}(c) and (d), 
predictions at 14 TeV by using the parameters
of the QCD-inspired scenario (labelled 3B in our notation) are given
for $p_2 = 0.77$ as in the GeV region and for $p_2$ logarithmically
decreasing with c.m.\ energy in the TeV region ($p_2 = 0.72$ at 14
TeV). They confirm the general trend of $\nbar_B(n_F)$ vs.\ $n_F$
seen at 900 GeV. The previously discussed scenarios 1B and 2B lead in
this case to very small modification (not shown).

Some comments are needed in order to understand the 
theoretical predictions of the weighted superposition
mechanism of the soft and semi-hard components for the general trend
of $\nbar_B(n_F)$ vs.\ $n_F$ and its linear behaviour found
experimentally in $|\eta|<4$ at 900 GeV by UA5 Collaboration
(Fig.~\ref{fig:MC}(b)).

The theoretical predictions are based on a separation of the two
components which comes from fits \cite{Fug} to the total charged MD
and not from an event-by-event classification.
The idea of separating the total class of events into two components
is correct in principle---as we have shown in our work---but is
questionable here in its application.
Different sets of NB parameters for the two components could lead to
good fits of the total charged MD. Two sets we used are taken from
\cite{Fug} and their predictions collected in bands in
Fig.~\ref{fig:MC}(b).    Results should be considered
therefore only indicative of the general trend, which we consider quite
satisfactory, especially for large $n_F$ values in the semi-hard
component and small $n_F$ values in the soft one, as visible in the
figure.

The second warning comes from intrinsic limitations of our approach.
The choice to perform our study analytically led us to consider the $p_1$
and $p_2$ parameters approximately constant throughout the GeV region,
an approximation which might turn out to be not correct and in any
case to imply small modifications of our results.

In addition our predictions are based on the separation of the events 
into two classes.
The onset of a third class of the same kind as that which could modify
the $b$ bending effect described in Fig.~\ref{fig:b}---when
properly inserted in our approach---could lead to different results.

All these considerations apply of course also to the FB multiplicity
correlation strength energy dependence discussed in Fig.~\ref{fig:b}
and testify that our predictions, in view of the lack of detailed
experimental data,  should be considerd only indicative of the
expected general trends.

\section{Conclusions}
General formulas for FB multiplicity correlation strength in the
framework of the superposition mechanism  of two weighted MD's
for different classes of events  as  functions of the average charged
particle multiplicities and dispersions of each class of events are given.
Assuming NB  regularity behaviour for 2- and 3-jet samples of events
in \ee\ annihilation at LEP energy, results obtained by OPAL
collaborations are correctly reproduced within experimental errors
in terms of the pure superposition effect, being FB multiplicity
correlation strengths  for the two separate classes of events negligible.

The same approach has been successfully extended to $pp$ collisions.
Differently from \ee\ annihilation, the FB multiplicity correlation
strengths in the two substructures (soft and semi-hard events) turn
out to be quite important and to lead to interesting predictions in
the TeV region in the three scenarios discussed by the Authors in a
previous paper and based on extrapolations of $pp$ collisions
properties in the GeV region.

In particular an interesting connection is found  between the
particle populations within clans, particle leakage from clans
in one hemisphere to the opposite hemisphere and superposition effect
between different substructure of the collision. This finding
favours structures with larger particle populations per clans and
the decrease of the average number of clans.  The FB charged particle
multiplicity correlation strength is predicted to bend in all
scenarios  in the TeV region.

The effects of the main assumptions of the present approach on the linear
relation between the average number of particles
emitted in one hemisphere as function of particle emitted in the
opposite hemisphere have also been studied and their expected general
trend at LHC explored.

\bibliographystyle{prstyR}  
\bibliography{abbrevs,bibliography}

\end{document}